\documentclass[10pt]{article}

\usepackage{graphicx}

\def\cites#1#2{\cite{#1}--\cite{#2}}
\def\eqref#1{(\ref{#1})}

\begin{document}

\noindent
{\large\bf Growth and Fluctuations of Personal Income}

\noindent
Yoshi Fujiwara$^*$,
Wataru Souma$^\dag$,\\
Hideaki Aoyama$^\ddag$,
Taisei Kaizoji$^{||}$ \&
Masanao Aoki$^\P$

\noindent{\small
  $^*$Keihanna Center, Communications Research Laboratory,
  Kyoto 619-0289, Japan,
  $^\dag$ATR Human Information Science Laboratories,
  Kyoto 619-0288, Japan,
  $^\ddag$Faculty of Integrated Human Studies, Kyoto University,
  Kyoto 606-8501, Japan,
  $^{||}$Division of Social Sciences, International Christian
  University, Tokyo 181-8585, Japan,
  $^\P$Department of Economics, University of California, Los Angels,
  90095-1477 USA
}

\vskip 0.5cm

{\bf 
  Pareto's law \cite{Pareto} states that the distribution of personal
  income obeys a power-law in the high-income range, and has been
  supported by international observations
  \cites{Gini}{Yakov01}. Researchers have proposed models
  \cites{Gibrat}{Solomon01} over a century since its
  discovery. However, the dynamical nature of personal income has been
  little studied hitherto, mostly due to the lack of empirical work.
  Here we report the first such study, an examination of the
  fluctuations in personal income of about 80,000 high-income
  taxpayers in Japan for two consecutive years, 1997 and 1998, when
  the economy was relatively stable. We find that the distribution of
  the growth rate in one year is independent of income in the previous
  year. This fact, combined with an approximate time-reversal
  symmetry, leads to the Pareto law, thereby explaining it as a
  consequence of a stable economy. We also derive a scaling relation
  between positive and negative growth rates, and show good agreement with
  the data. These findings provide the direct observation of the
  dynamical process of personal income flow not yet studied as much as
  for companies
  \cites{SIMON77}{Axtell01}.
}

\vskip 1cm

Flow and stock are the fundamental concepts in economics. They refer
to a certain economic quantity in a given period of time and its
accumulation at a point of time respectively. Personal income and
wealth can be regarded as flow and stock observed at each
individual in a giant dynamical network of people, which is open to
various economic activities. The Italian social economist Vilfredo
Pareto \cite{Pareto}, more than a century ago, studied the distribution
of personal income and wealth in society as a characterization of a
country's economic status. He found that the high-income distribution
follows a power-law: the probability that a given individual has
income equal to, or greater than $x$, denoted by $P_>(x)$, obeys
\begin{equation}
  P_>(x) \propto x^{-\mu} ,
\label{eq:pareto}
\end{equation}
with a constant $\mu$ called Pareto index. This phenomenon, now known
as a classic example of fractals, has been observed
\cites{Gini}{Souma01} in many different countries, where $\mu$ varies
typically around 2 reflecting economic conditions.

Recent high-quality digitized data proves that the law holds for
high-income range often with remarkable accuracy, and allows precise
estimate of Pareto index over years. Fig.1a shows the distribution
of Japanese personal income in the year 2000, derived from available
data of the Japanese National Tax Administration (NTA) (corresponding
to UK Board of Inland Revenue). Power-law behavior is a salient
feature characterizing high income range nearly three orders of
magnitude. 

Understanding the origin of the law has importance in economics
because of linkage with consumption, business cycle and other
macro-economic activities, and also for practices in assessment of
economic inequality \cite{Champer96}. Many researchers, recently
including those in non-equilibrium statistical physics, have proposed
models \cites{Gibrat}{Solomon01}. Some theories were based on
multiplicative stochastic processes. A classic theory by
Gibrat \cite{Gibrat} assumed that personal income depends on a number of
causes each of which has a proportional effect that is independent of
the proportional effects of the others, and also of initial income (law
of proportionate effect). This theory, basically a random walk in
logarithmic scale of income, predicts log-normal distribution of income
with Gaussian growth rate, both in disagreement with actual data for
high income. One could introduce to the process a boundary effect that
income should not be less than a value, and derived a power-law
distribution \cite{Champer53,Mandelbrot61}. Another approach is to
construct a simple but minimal economic model in a network of
wealth \cite{Bouchaud00}. Actually there have been proposed many kinds of
scenarios \cite{SORNETTE00} which predict a power-law distribution as a
static snapshot. However, in order to test models, it has been highly
desirable to have direct observation of the dynamical process of growth
and fluctuations of personal income.

For that purpose, we employ Japanese income tax data which covers most
of the power-law region in Fig.1a. It is an exhaustive list of all
taxpayers with full names, addresses and tax amounts, who paid 10
million yen or more in a year through tax offices of the Japanese
National Tax Administration (NTA). The data were gathered from all the
NTA offices. In Fig.1b, Pareto indices, estimated from such income
tax data, since 1987 to 2000, are plotted. $\mu$ changes annually
around 2 with an abrupt jump between 1991 and 1992.  Before the years,
Japanese economy experienced abnormal rise of prices in the risky
assets of lands and shares due to speculative investment (``bubble''),
after which those prices fell rapidly. We examined a relatively stable
period in economy, namely 1997 and 1998. The complete datasets of
93,394 persons in 1997 and 84,571 in 1998 were used. Identification of
individuals who are listed in both of the years were done if and only
if his/her full name uniquely and exactly matches in both years with
the same address (zip-code). Duplicate matches were only a few cases
that were discarded. We assumed that the change of address and name is
negligible in fraction. The number of the common set of those
appearing in the two consecutive years was 52,902. The rest of persons
in 1997 and 1998 can be therefore regarded as those disappearing from
or novel in the list.

The common set is shown by the scatter plot in Fig.2, where each
point represents a person who paid income tax of $T_1$ in 1997 and $T_2$
in 1998 (both in units of thousand yen). This represents the joint
distribution $P_{12}(T_1,T_2)$. The plot is consistent with approximate
time-reversal symmetry in the sense that the joint distribution is
invariant under the exchange of the values $T_1$ and $T_2$.

Now the quantity of our concern is the annual change of individual
income-tax, or growth. Growth rate is defined as $R=T_2/T_1$. It is
customary to use the logarithm of $R$, $r\equiv\log_{10}R$. We examine
the probability density for the growth rate $P(r|T_1)$ conditioned that
the income $T_1$ in the initial year is fixed. The result is shown in
Fig.3. Here we divide the range of $T_1$ into logarithmically equal
bins as $T_1\in[10^{4+0.2(n-1)},10^{4+0.2n}]$ with $n=1,\cdots,5$. For
each bin, the probability density for $r$ was calculated. As shown in
the figure, different plots for $n$ collapse onto each other. This fact
means that the distribution for the growth rate $r$ is statistically
independent of the initial value of $T_1$.  In a mathematical notation,
we found that
\begin{equation}
  P_{1R}(T_1,R)=P_1(T_1)\,P_R(R) ,
\label{eq:indep}
\end{equation}
where $P_{1R}$ is the joint distribution for $T_1$ and $R$, $P_1$ and
$P_R$ are the distributions for $T_1$ and $R$ respectively.

This ``universal'' distribution for the growth rate has a skewed and
heavy-tailed shape with a peak at $R=1$. How is such a functional form
consistent with the approximate time-reversal symmetry shown in Fig.2? 
The answer to this question leads us to an important bridge from
the fluctuations of growth rates to the Pareto law as follows. The
time-reversal symmetry (Fig.2) claims that
$P_{12}(T_1,T_2)=P_{12}(T_2,T_1)$. One can easily see that under the
variable transformation from $(T_1,T_2)$ to $(T_1,R)$, the equality
$P_{1R}(T_1,R)=T_1 P(T_1,T_2)$ holds. This equality, together with the
time-reversal symmetry and the statistical independence of equation
\eqref{eq:indep}, leads us to the relation:
\begin{equation}
  P_1(T_2)/P_1(T_1)=R\,P_R(R)/P_R(1/R) .
\label{eq:rel}
\end{equation}
The left-hand side is a function of $T_1$ and $T_2$, while the
right-hand side is a function of the ratio $R$ only. We can then
conclude that the distribution $P_1$ obeys a power-law: $P_1(x)\propto
x^{-(\mu+1)}$, whose integral form gives the expression, equation
\eqref{eq:pareto}. Thus the independence in the growth rate of the past
value and the time-reversal symmetry requires the Pareto law.

In addition, we have a scaling relation following immediately from the
above relation \eqref{eq:rel} and equation \eqref{eq:pareto}:
\begin{equation}
  P_R(R)=R^{-(\mu+2)}\,P_R(1/R) .
\label{eq:scaling}
\end{equation}
This equation relates the positive and negative growth rates through
the Pareto index $\mu$. In Fig.3, we fitted $P_R(R)$ for the region
of positive growth $r>0$ with an analytic function, and then plotted
its counter part for negative growth rate $r<0$ derived from the
scaling relation, equation.\eqref{eq:scaling}. The result fits the
data in the region quite satisfactorily.

In summary, the statistical independence of growth rate, the
approximate time-reversal symmetry and the power-law are consistent
with each other.  According to a sample survey by NTA on income
earners with total income exceeding 50 million yen and on sources of
earning, their sources are employment income, income from real estate,
capital gains from lands and shares. In fraction of income amount,
capital gains from risky assets considerably exceed than other
non-risky income sources. It would be expected that asymmetric
behavior of price fluctuations in those risky assets and accompanying
increase of high-income persons causes breakdown of time-reversal
symmetry, which necessarily brings about the invalidity of Pareto's
law. This was actually the case in the ``bubble'' phase of Japanese
economy, during which the prices of risky assets, especially of lands,
rise abnormally compared to their fundamental values. Fig.4 shows the
cumulative distributions of income tax in 1991 (peak of speculative
bubble) and 1992. One can observe that the 1991 data cannot be fitted
by the Pareto's law in the entire range of high-income, compared to
the 1992 data.

Our finding in this work shall serve as an empirical test for models of
personal income and wealth, where people make choice among assets with
different risks and returns, with changing degrees of freedom. Personal
income is not a single example of such systems but other systems
comprised by economic agents \cite{AOKI02} including companies,
institutions and nations might be worth being examined from a new
look. Indeed comparison with and similar analysis in company growth,
which has been studied extensively \cites{SIMON77}{Axtell01}, would
be an interesting subject, where the Zipf law ($\mu=1$) is observed.

\section*{Acknowledgements}

We thank M.~Ojima and Teikoku Databank, Ltd. for kindly providing
high-quality data. This work was supported by grants from the
Telecommunications Advancement Organization in Japan.

\clearpage

\thispagestyle{empty}

\begin{figure}
  \centering
  \includegraphics[width=.8\columnwidth]{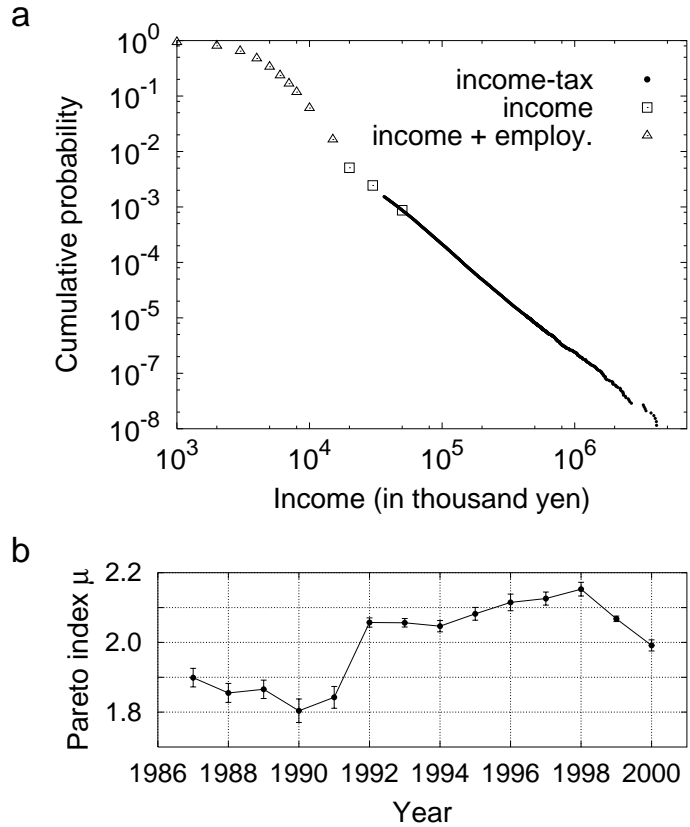}
  \caption[]{%
    Personal income in Japan. %
    a. Cumulative probability distribution of personal income from low
 to high income range in the year 2000. A
    data-point represents the probability (vertical axis) that a
    person has income equal to or more than the income of the
    horizontal value.  Three datasets available from the Japanese
    National Tax Administration (NTA) were used. (i) Income tax data
    (dots) is the exhaustive list of all taxpayers, about 80,000, who
    paid income tax of 10 million yen or more. Tax value was converted
    to income uniformly by the same proportionality following the
    previous work\cite{Aoyama00}. (ii) Income data (squares), a
    coarsely tabulated data for all the persons, about 7,273,000, who
    filed tax return. (iii) Employment income data, a sample survey
    for the salaried workers in private enterprises, about 44,940,000.
    Under the Japanese taxation, all persons with income exceeding 20
    million yen have obligation to file final declaration to the NTA
    in each year. Thus the dataset (ii) includes all the
    persons listed in (i), so we have a reliable profile in the high
    income range ($>$ 20 million yen).  For lower income, upper-bound
    estimate (triangles) was given by overlapping the datasets
    (ii) and (iii) which was found relatively good\cite{Souma01}.\\
    b. Annual change of Pareto index $\mu$ from the year 1987 to 2000.
    The complete list of income tax data in each year was used.
    Excluding top 0.1 percent and bottom 10 percent, samples equally
    spaced in logarithm of rank were plotted, from which slopes were
    estimated by least-square-fit. Error bars shown are standard error
    (90\% level) of the estimate $\mu$ (dots).%
  }
  \label{fig:pareto}
\end{figure}
\thispagestyle{empty}

\begin{figure}
  \centering
  \includegraphics[width=.8\columnwidth]{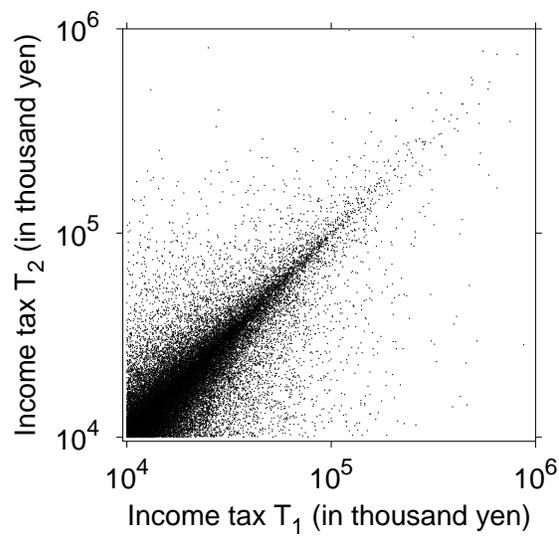}
  \caption[]{%
    Scatter-plot of all the individuals whose income tax exceeds 10
    million yen both in the years 1997 and 1998. These points (52,902)
    were identified from the complete list of high-income taxpayers in
    1997 (93,394) and in 1998 (84,571) (numbers in parentheses), with
    income taxes $T_1$ and $T_2$ in each year. A few points with $T_1$
    and/or $T_2$ exceeding $10^4$ exist but are not shown here.%
  }
  \label{fig:joint}
\end{figure}
\thispagestyle{empty}

\begin{figure}
  \centering
  \includegraphics[width=.8\columnwidth]{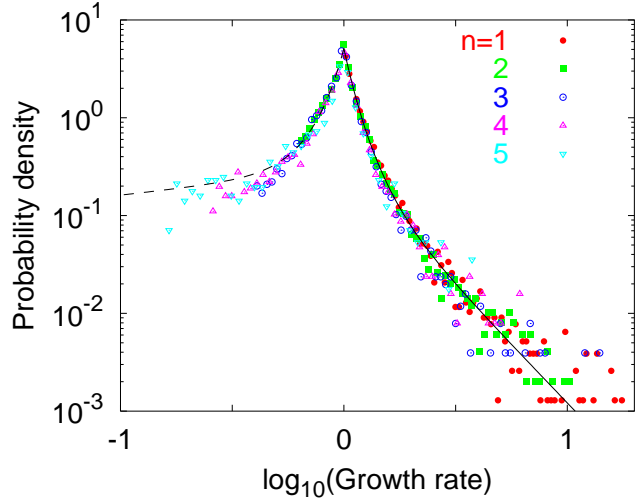}
  \caption[]{%
    Probability density $P(r|T_1)$ of the growth rate
    $r\equiv\log_{10}(T_2/T_1)$ from year 1997 to 1998. Note that due
    to the limit $T_1 > 10^4$ (in thousand yen), the data for large
    negative growth, $r<4-\log_{10}T_1$, are not available. Different
    bins of initial income-tax with equal size in logarithmic scale
    were taken as $T_1\in[10^{4+0.2(n-1)},10^{4+0.2n}]$
    ($n=1,\cdots,5$) to plot probability densities separately for
    each such bins. All the densities collapse upon a same curve. This
    fact means that $P(r|T_1)$ does not depend on $T_1$. The solid
    line in the portion of positive growth ($r>0$) is an analytic fit.
    The dashed line ($r<0$), on the other side, is calculated from the
    fit by the predicted relation given in equation
    \protect{\ref{eq:scaling}}, which follows from the statistical
    independence shown here and approximate time-reversal symmetry.
    The predicted density of negative growth fits quite well with the
    actual data.%
  }
  \label{fig:gr}
\end{figure}
\begin{figure}
  \centering
  \includegraphics[width=.8\columnwidth]{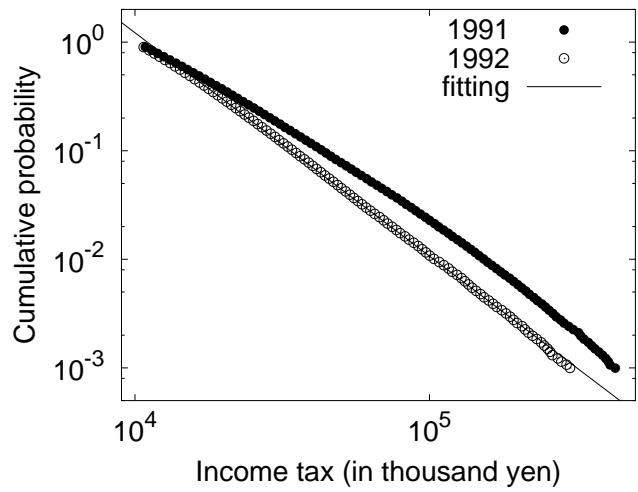}
  \caption[]{%
    Cumulative probability distributions of income tax in 1991 and
    1992. The Pareto index for 1992 data was estimated by excluding
    top 0.1 percent and bottom 10 percent, sampling equally in
    logarithmic scale, and estimating by least-square-fit, which is
    the fitted line ($\mu=2.057\pm 0.005$).%
  }
  \label{fig:bubble}
\end{figure}

\end{document}